\documentclass[preprint,3p]{elsarticle} 
\usepackage{amssymb}
\newcommand{\degree}{\ensuremath{^\circ}} 
\newcommand{\microns}{$\mu$m }
\newcommand{\micron}{$\mu$m}
\newcommand\arcsec{\mbox{$^{\prime\prime}$}}%

\usepackage{natbib}
\bibpunct{(}{)}{;}{a}{}{,}

\journal{icarus}

\begin{document} 

\begin{frontmatter}
\title{Detectability of Planetesimal Impacts on Giant Exoplanets}

\author[nau,ciw]{Laura Flagg\corref{cor1}}
\ead{laura@nau.edu}

\author[ciw]{Alycia J. Weinberger}
\ead{weinberger@dtm.ciw.edu}
\author[ct]{Keith Matthews}

\address[nau]{Department of Physics and Astronomy, Northern Arizona University, Flagstaff, AZ 86011-6010, USA}
\address[ciw]{Department of Terrestrial Magnetism, Carnegie Institution of
Washington, 5241 Broad Branch Road NW, Washington, DC 20015, USA}
\address[ct]{Caltech Optical Observatories, California Institute of Technology, MC 301-17, Pasadena, CA 91125, USA}
 \cortext[cor1]{Corresponding author}
      
\begin{abstract} 

The detectability of planetesimal impacts on imaged exoplanets can be
measured using Jupiter during the 1994 comet Shoemaker-Levy 9 events as a
proxy. By integrating the whole planet flux with and without impact spots, the
effect of the impacts at wavelengths from 2 - 4 $\mu$m is revealed. Jupiter's reflected light spectrum in the
near-infrared is dominated by its methane opacity including a deep band at 2.3
\micron.  After the impact, sunlight that would have normally been absorbed by the large amount of methane in Jupiter's atmosphere was instead reflected by the cometary material from the impacts.  As a result, at
2.3 \micron, where the planet would normally have low reflectivity, it
brightened substantially and stayed brighter for at least a month.
\end{abstract} 

\end{frontmatter}

\section{Introduction}
The frequencies with which giant planets and cold circumstellar debris disks
encircle old (i.e. $>$1 Gyr) Sun-like stars are both $\sim$20\%
\citep{eir13,marshall2014}.  The debris disks are generated in the
collisions and evaporation of planetesimals, analogous to the comets and
asteroids of today's Solar System. That most debris disks contain cold,
$\sim$50 K, dust indicates that the reservoirs of parent bodies are at $\sim$100
AU from the parent stars, in a region analogous to our Edgeworth-Kuiper Belt. These cold
debris disks appear to be equally common for planet-hosting and non-planet
hosting stars \citep{bry09}, though the known planets in these systems are all within a
$\sim$1 AU of their stars.  Detectable disks are at least 10--100$\times$ dustier than
our Edgeworth-Kuiper Belt \citep{bry06,eir13}, with correspondingly larger parent body
populations and are thus even more subject to frequent collisions between
bodies.

In today's Solar System, planetesimals infrequently hit planets; although, in
the past, such impacts were much more common. The cratering rates on the Moon
and Mars indicate that the impact rates were high prior to $\sim$800 Myr after
the formation of the Solar System, perhaps ending in a ``Late Heavy Bombardment''
\citep[see e.g.][and references therein]{str05}.  Today, a comet larger than 1 km in diameter impacts Jupiter
only every few hundred years \citep{nak98,zah03,sch04}. One such collision was that of Jupiter and Shoemaker-Levy 9 (SL9) in July 1994.  SL9 was a Jupiter Family comet, a class of objects
believed to originate in the Edgeworth-Kuiper Belt \citep{sch04}.  During a pass of
Jupiter in 1992  the comet was tidally disrupted and broke into $\sim$25
fragments \citep{sch04}.  Three of these fragments were
still quite large -- $>$1 km in diameter \citep{har04,cra97}.  Fifteen fragments produced impacts detected by the Hubble Space Telescope \citep{ham95}, as the comet impacted the planet over the course of seven days.

In this paper, the impact of Shoemaker-Levy 9 on Jupiter is considered as a
case study for assessing the observability of impacts on giant
exoplanets. The comet impact caused visible changes in Jupiter's atmosphere in
two major ways on two different timescales.  The immediate impacts were very
bright as the bolides entered the Jovian atmosphere, but these effects lasted
only seconds. Afterward a brightening due to thermal emission occurred that
lasted on the order of several hours \citep{nic95}. These short duration phenomena are unlikely to be
observed unless the timing of the comet impact is known in advance, as for SL9, or a
planet were being continuously monitored.  The second major effect of SL9 was
to deposit high-altitude particulates that blocked sunlight from reaching the
deeper atmosphere. The smallest particulates ($<$0.5 \micron) stayed at those altitudes for at least two years \citep{san98}.


Jupiter's atmosphere contains 0.2\% of
methane, which has deep absorption bands in the near-infrared, especially near
2.3 \micron\ in the K-band and 3.3 \micron\ in the L-band \citep{tay84}.  When
SL9 debris at higher altitudes reflected photons from the sun and prevented them from being absorbed by the methane at lower altitudes, 
Jupiter appeared brighter near the impact spots long after the initial effects
of the thermal heating had ended \citep{ban96}. Because of the tidal disruption, the
fragments hit at a wide range of Jovian
longitudes, efficiently covering the atmosphere at  $\sim$45$\degree$ South latitude.

For comparable effects to occur with impacts of comets and exoplanets, those planets must have a source of opacity in their atmosphere, like methane, whose absorption could be blocked by high-altitude particulates.  The results from this paper should be applicable  to exoplanets as current data and models indicate that exoplanets will have such a source of opacity \citep{mor14}.  
\section{Observations}

Near-infrared observations of Jupiter were obtained at the Hale 200-inch
Telescope at Palomar Observatory on 1994 July 22-27 UT, i.e. the nights
following the last of the fragment impacts on 22 July. Followup observations
were obtained over the next month on 16-19 August and 27-29 August.  The only
photometric nights were July 22-23, July 25-26, and August 28-29 UT.

Images were taken with the D78 Cassegrain camera, which employed a
256$\times$256 HgCdTe (NICMOS3) array with a pixel size of 0.125\arcsec\ square,
for a field-of-view of 32$\times$32 arcsec.  The camera was rotated to put the
North-South axis of Jupiter along the detector columns. Over the $\sim$1 month of observations, Jupiter's apparent
diameter ranged from approximately 37.7 to 34.2\arcsec\ (radius of 151 to 137
pixels) along the equatorial axis and from 35.2 to 32\arcsec\ (radius of 141
to 128 pixels) along the polar axis.  More of the planet could be seen in the
late August images than in the July images, as Jupiter was moving away from
Earth during that time, going from a distance of about 5.24 AU to 5.78 AU. In
addition, to avoid cosmetic defects on the array --- dead pixels which would have prevented photons from the south pole from being counted --- the planet was generally
placed somewhat higher than center. Therefore, Jupiter was too large to
completely fit on the detector, and a portion of the top of the planet was cut
off in every image.  The detector also had many bad pixels, but fortunately
most of them were in the top half of the array and not near where the impact
spots appeared.  These pixels were replaced with the weighted average flux of surrounding pixels, which had minimal impact as the total planetary flux was calculated and not the detailed distribution of flux. Through July 25, a second instrument, Spectro-Cam 10, was mounted
simultaneously with D78, and its pick-off mirror vignetted on the top of the
D78 images.  On July 26, condensation from its dewar dropped water onto D78, causing
bright spots on the images.  Again, while this contaminated the quality of the top part of the image, the lower part, with the impact spots, was unaffected. 

The observations are summarized in Table \ref{tab:obslog}. Sequences of images
of Jupiter were collected across the K and L-bands.  A typical sequence
started with an image in a methane filter centered at 2.276 \microns with a
bandwidth of 0.17 \micron.  The sequence continued with images
taken in 1\% bandwidth continuously variable filter (CVF) at eight central
wavelengths in the K-band (2.00, 2.05, 2.10, 2.14, 2.17, 2.20, 2.25, and 2.35
\micron). At the end of the K-band CVF sequence, another methane filter image
was taken.  The sequences finished with L-band CVF observations at 11 central
wavelengths (3.00-4.00 \micron\ at 0.10 \micron\ intervals). In both bands,
individual frame integration times were either 1 or 0.4 seconds, repeated and
co-added to reach at least 10 seconds total integration time at each filter.  
In each sequence, the telescope was nodded to sky for an image after every
image of Jupiter and these sky images were used for background
subtraction.
\begin{table}[htbp]

  \centering
      \begin{tabular}{rrrrrr}
    date  & times & wavelengths & int time & airmass range & longitude range \\
 & UT & \micron & sec & &degrees\\
    \hline
    July 22 & 23:14 to 23:17 & 2.276 & 10   & 2     & 1-2 \\
    July 22 & 23:18 to 23:30 & 2.0-2.35 & 20     & 1.9-2 & 3-10 \\
    July 22 & 23:37 to 23:59 & 3.0-4.0 & 40   & 1.7-1.9 & 14-27 \\
    July 23 & 00:00 to 00:22 & 3.0-4.0 & 40     & 1.6-1.7 & 28-40 \\
    July 23 & 02:14 to 02:16 & 2.276 & 1     & 1.4   & 114 \\
    July 23 & 02:27 to 02:47 & 3.0-4.0 & 40   & 1.4-1.5 & 117-130 \\
    July 23 & 02:54 to 02:56 & 2.276 & 20     & 1.5   & 134 \\
    July 23 & 02:56 to 03:04 & 2.0-2.35 & 20     & 1.5   & 135-140 \\
    July 23 & 03:06 to 03:10 & 2.276 & 20     & 1.5   & 141 \\
    July 25 & 23:30 to 23:35 & 2.276 & 10/20   & 1.8   & 102 \\
    July 25 & 23:35 to 23:46 & 2.0-2.35 & 20     & 1.7-1.8 & 105-111 \\
   July 25 & 23:47 to 23:48 & 2.276 & 10     & 1.7 & 112 \\
    July 25-26 & 23:49 to 00:23 & 3.0-4.0 & 80   & 1.5-1.7 & 113-133 \\
    July 26 & 00:41 to 01:16 & 3.0-4.0 & 80     & 1.4-1.5 & 145-166 \\
    July 26 & 01:30 to 01:32 & 2.276 & 0.8   & 1.4   & 174 \\
    July 26 & 02:18 to 02:19 & 2.276 & 20     & 1.4   & 203 \\
    August 28 & 23:02 to 23:23 & 3.0-4.0 & 40   & 1.5   & 177-187 \\
    August 28 & 23:24 to 23:25 & 2.276 & 20   & 1.5   & 191 \\
    August 28 & 23:26 to 23:36 & 2.0-2.35 & 20     & 1.5   & 192-198 \\
    August 28 & 23:37 to 23:42 & 2.276 & 10/20   & 1.2-1.5 & 199 \\
    August 29 & 00:45 to 00:56 & 2.276 & 10   & 1.5   & 246 \\
    August 29 & 00:57 to 01:07 & 2.0-2.35 & 20     & 1.6   & 247-253 \\
    August 29 & 01:08 to 01:09 & 2.276 & 20   & 1.6   & 254 \\
    August 29 & 01:09 to 01:45 & 3.0-4.0 & 80     & 1.6 to 1.7 & 255-275 \\
    August 29 & 01:47 to 01:48 & 2.276 & 20   & 1.7   & 278 \\
    August 29 & 01:49 to 01:59 & 2.0-2.35 & 20     & 1.7-1.8 & 278-284 \\
    August 29 & 02:00 to 02:01 & 2.276 & 20   & 1.8   & 285 \\
    August 29 & 02:06 to 02:21 & 3.0-3.7 & 40   & 1.9-2.0 & 289-297 \\
    \end{tabular}%
   \caption{Sequences of images of Jupiter taken at the given dates, times, and wavelengths.   \label{tab:obslog}}
\end{table}%

Generally, two sequences were completed each night. Each image sequence lasted
20 minutes.  During this time Jupiter rotated $\sim$12 deg.  
The 2 \micron\
data were previously used in \citet{ban96} for a study of the aerosol
clouds formed by the impacts and the Jovian winds that dissipated them.

\begin{table}[htbp]
  \centering  
    \begin{tabular}{rrr}
    star  & K band magnitude & L band magnitude \\
    \hline
    HD 129655 & 6.69  & 6.66 \\
     HR 5107 & 3.06 & 3.05 \\
   HR 6707 & 3.20 & 3.15 \\
    HR 6147 & 2.26 & 2.18 \\
    HD 161903 & 7.02 & 6.98 \\
    \end{tabular}%
  \label{tab:mag}%
\caption{The standard star magnitudes for HD 129655, HR 6707, and HD 161903 were adopted from  \citet{eli82} and HR 5107 and HR 6147 from CIT unpublished  measurements (G. Neugebauer, personal communication).   }
\end{table}%

Standard stars with known magnitudes (Table \ref{tab:mag})  were observed on the three photometric nights, as summarized in
Table \ref{tab:stdlog}.  These bright stars were nodded on the detector for
sky subtraction.  HR 5107 was streaked with the chopping secondary to avoid
saturation.

\begin{table}[htbp]
  \centering  
    \begin{tabular}{rrrr}
    date  & times (UT) & stars & airmass range \\
    \hline
    July 23 & 03:11 to 03:43 & HD 129655 & 1.3 \\
    July 26 & 01:33 to 02:07 & HD 129655, HR 5107 & 1.2-1.3 \\
    August 28-29 & 23:57 to 00:39 & HD 129655, HR 6707, HR 6147, HD 161903 & 1.2-1.7 \\
    \end{tabular}%
\caption{Standard stars used on the given nights.    \label{tab:stdlog} }
\end{table}%

\section{Data Analysis}

\subsection{Image Processing}

Basic imaging processing steps were: a linearity correction for the response
of each pixel, sky subtraction using the offset sky image, and flat-fielding
to correct for pixel-to-pixel response variations.  The flat-field image was
created by dark subtracting average sky images  at 2.27 $\mu$m and 3.5 $\mu$m for the K-band and L-band observations respectively.

A large number of bad and intermittently bad pixels had to be identified and corrected. To
determine the location of these pixels, about 50 images with high signal-to-noise ratios
were randomly chosen. In each image, obviously bad pixels, those with values
over two orders of magnitude different from any surrounding pixels, were
identified first.  Then, pixels off by one order of magnitude or more were
added to a second list in order to cross check.  If a pixel was
listed for 75\% of the files, it was included in a general bad pixel mask.  To determine which pixels were bad for
each specific image, a program was created that went through each file and
determined which pixels were not within two orders of magnitude of the median
of the surrounding pixels.  This was combined with the general bad pixel mask
to create a mask for each individual file.

Sky subtraction using the offset sky images was imperfect due to the time
difference between the image of Jupiter and its corresponding blank sky
image. An additional sky correction was made by finding an average value
of the pixels outside the planet by at least 10 pixels.  Not included were
pixels contaminated or vignetted or marked as a bad pixel by the image's bad
pixel mask.  After an initial sky average was calculated, any pixel whose
value was not within 3 sigma of this average was excluded, and a second
average was calculated.  This second average was what was subtracted from the
image.

Each pixel marked as a bad pixel needed to be corrected.  Pixels within three
pixels of the bad one are used to calculate a new value based on their average, with pixels closer to
the bad pixel being weighted more than pixels further away.  Other bad pixels
within the three pixel radius are excluded.

\subsection{Photometric Calibration}

The standard star magnitudes for HD 129655, HR 6707, and HD 16190 were adopted from  \citet{eli82} and HR 5107 and HR 6147 from CIT unpublished  measurements (G. Neugebauer, personal communication)  
(see Table \ref{tab:mag}). Standard stars were observed at airmasses of
$\sim$ 1-1.8.

Basic image processing for the standard star frames was the same as described
in the previous section. In each of the two
standard star images, a 2D Gaussian fit was used to
determine the center of the star.  Aperture photometry was done in a 20 pixel
radius except for the elongated images of HR 5107, in which a rectangular mask
was used. The aperture was large enough to encompass all of the
stellar flux. 


The airmass correction was calculated using  the standard airmass correction for Palomar.

\subsection{Whole Planet Photometry}

In order to compute the total brightness of Jupiter at each wavelength and
time, we had to identify the pixels belonging to the planet in each image.
Because Jupiter bulges at the equator, its shape was treated as an ellipse
with values of its semi-major and semi-minor axes required. The equatorial
radius of Jupiter is 71492$\pm$4 km, and the polar radius is 66854$\pm$10 km \citep{sei07}. As the uncertainties in the radii are negligible, they were ignored in the calculations. As observed,
the radius in pixels was:
\begin{displaymath}
r=\frac{206265}{0.125}arctan(\frac{R}{d})
\end{displaymath}
where $R$ is the radius of Jupiter in km, $d$ is the distance from Earth to
Jupiter in km, 206265 is the number of arcseconds in one radian, and 0.125 is
the pixel scale of the camera.

An ellipse with those measurements was manually aligned with each image, and
the center of the ellipse taken as the center of the planet. 

\begin{figure}
\centering
\includegraphics[width=3.2in]{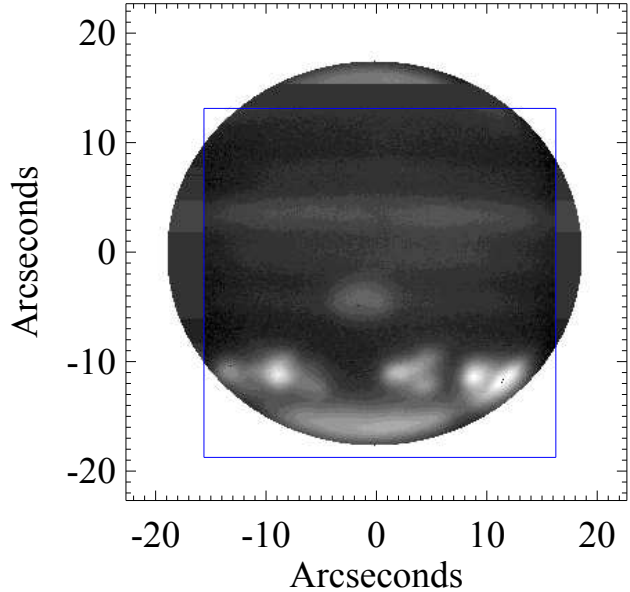}
\caption{An example of a  recreation of Jupiter, using an image taken on July 25 at 2.27 \microns. The blue box indicates what was in the original image.  Everything outside of the blue box had to be extrapolated.  The sides that were cut off were extrapolated by averaging the flux for that latitude. The north pole was extrapolated by  duplicating the south pole but at half its flux.  The scaling in this image is the square root of the flux, so that the different latitudes can be more easily seen.\label{fig_repro}}
\end{figure}

The JPL HORIZONS
\citep{gio96}\footnote{http://ssd.jpl.nasa.gov/?horizons} ephemeris
was used to determine the central longitude of the planet and its radius at the time of each
image. The planet was divided into three regions based on latitude.  The first was
the region where the planet was impacted, which ranged from approximately 60$\degree$
south latitude to 30$\degree$ south latitude, the south polar region which is everything south of the spots region (i.e. 30$\degree$ S), and a region near the equator that was relatively symmetric with respect to longitude
from approximately 15$\degree$ south latitude to the equator.  

 Unobserved areas of the planet were extrapolated as follows. The average
pixel value in every detector row (latitude) sampling the planet was used to
extrapolate the image out to the ephemeris radius. For the north polar region, the south pole region was duplicated but at half its flux.  The flux was halved because the north polar region has less total brightness than the south polar region \citep{kim10}.  In each sequence of images, the flux from each region was
summed.

We did not have observations of Jupiter in the same filter sequence prior to the impacts.  However, we can construct a model of a pre-impact Jupiter-like planet, i.e. a gas giant planet with much of its flux coming from the poles, by combining two polar regions (i.e. duplicating the measured south polar region in the north) and a
central region, which is everything between 60$\degree$ north latitude and 60$\degree$ south latitude. The pixels in the central region were set at the average flux from around Jupiter's equatorial region between 0$\degree$ and 10$\degree$ north latitude.    A Jupiter-like planet after the SL9 impact consists of the
same two polar regions, a similar central region from 60$\degree$ north latitude to 30$\degree$ south latitude, and the impact spots from 30$\degree$ south latitude to 60$\degree$ south latitude.

\begin{figure}
\centering
\includegraphics [width=3.2in]{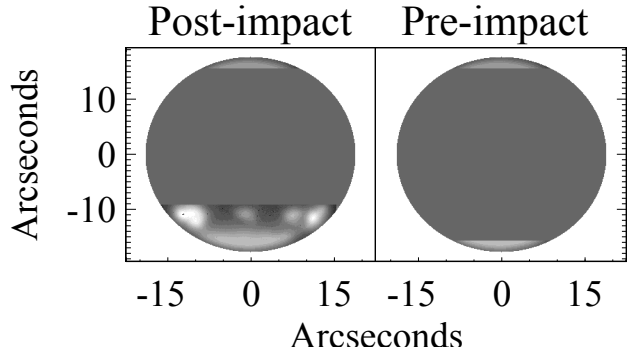}
\caption{An example of the model giant planets  created at 2.27 \microns.  These model planets were based on Jupiter with much of the flux coming from the poles.   They are identical in the two polar regions and the central region from 60$\degree$ north latitude to 30$\degree$ south latitude.  The pixels in the central region are set at the average flux from around Jupiter's equatorial region between 0$\degree$ and 10$\degree$ north latitude.   From  30$\degree$ south latitude to 60$\degree$ south latitude, the figure on the left has the post-impact planet with impact spots, while the figure on the right, representing the pre-impact planet, has an extended central region.
\label{fig_imagspotsnospots}}
\end{figure}

For each sequence of images, we measure the
surface brightness of the planet in each of the three regions and construct a simulated
planet.  The measured pixel values are converted to magnitude by taking their
logarithm multiplied by -2.5.  A specific airmass correction for each region or
planet was added to this magnitude.

\begin{displaymath}
AMC=z-(am-1.0)*amc
\end{displaymath}
where $z$ is the zeropoint for the wavelength, $am$ is the airmass for the image from which the planet or region was taken, and $amc$ is the general airmass correction for the wavelength.

The apparent planet albedo, i.e. the planetary radiance divided by the
incoming Solar flux, is given by:
\begin{displaymath}
\frac{I}{F}=\frac{\pi r^2}{\Omega}10^{0.4(K-m)} 
\end{displaymath}
where $r$ is the Jupiter-Sun distance in AU, $\Omega$ is the pixel solid
angle,  $K$ is the apparent magnitude of the Sun \citep[e.g.][]{klassenbell2003}.

\section{Results}

Exoplanets will not be spatially resolved, so total integrated planet
brightness would be the only measurable quantity.  Brightness is a function of the planet's distance from the star --- which can be estimated and corrected for --- and the planet's total albedo.  We examined how this total albedo would compare at three different times: prior to the impacts, the early days following the impacts, and a month after the impacts.

\begin{figure}
\centering
\includegraphics[width=6.4in]{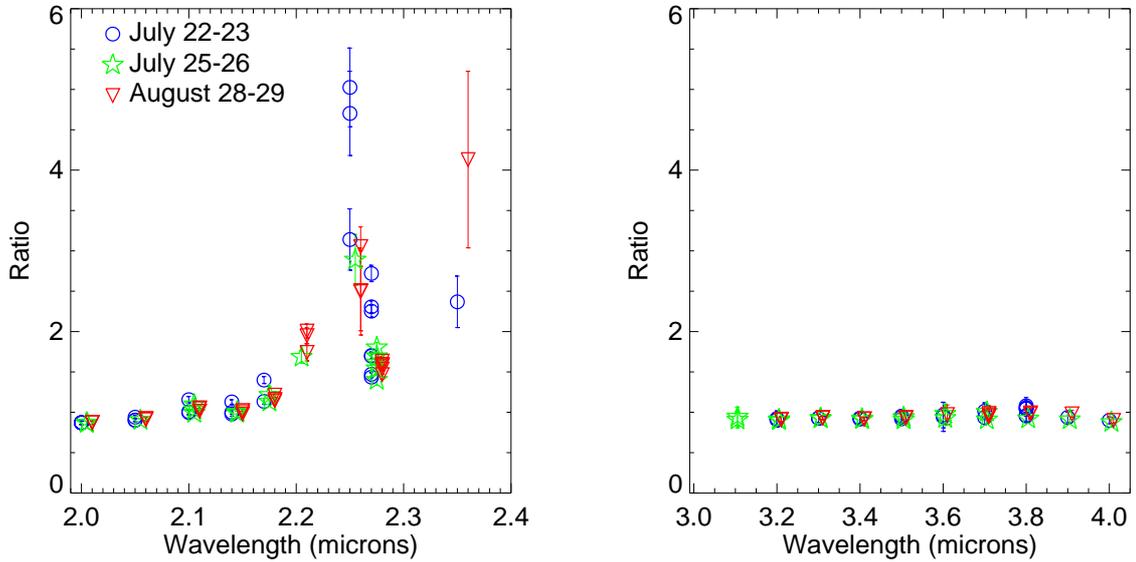}
\caption{(Left) The ratio of $I/F$ between the model planets with
  impact spots to the model planets without impact spots as a function of K-band wavelength.  At most wavelengths, the difference
between the two was negligible, resulting in a ratio near 1. However, near 2.3 \micron, where there is deep methane absorption, the change was much more noticeable.  (Right) Same ratios for the L-band. Despite a deep methane absorption band near 3.3 \micron, the ratio shows no change as a function of wavelength. Thermal emission leaking out of the planet apparently swamps the signal of the impact clouds.  The data from
  the different days are artificially off-set in wavelength for
  clarity. \label{fig_imaginaryplanetratio}}  
\end{figure}

\begin{figure}
\centering
\includegraphics[width=6.4in]{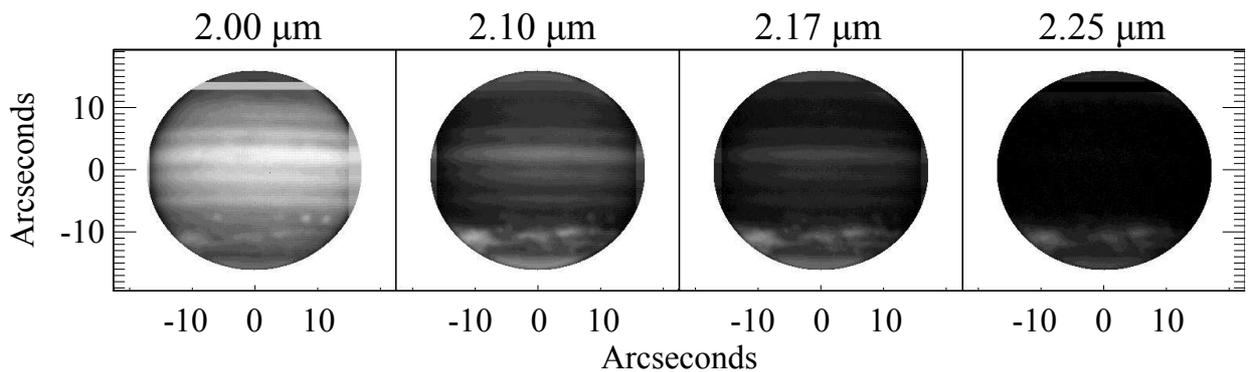}
\caption{Reproductions of Jupiter from August 29 at 2.0, 2.1, 2.17, and 2.25 \micron. At shorter wavelengths, the impact spots are more difficult to see, because the rest of the planet is so bright.  In the methane band, which is prominent at the longer wavelengths, the rest of the planet is fainter, so the spots are clearer.\label{fig_wlcompk_more}}  
\end{figure}

To compare what a planet with impact spots would look like to one without impact spots, the ratio of I/F from the model giant planets, such as the ones in Figure \ref{fig_imagspotsnospots}, was taken.  The impacts would be most easily detected at wavelengths where this ratio is very high.  The results are displayed in Figure \ref{fig_imaginaryplanetratio}.  At most wavelengths, the difference between the two was negligible. Wavelengths at which the planet is already very bright are those
without major sources of atmospheric opacity.   For example, in the K-band, the increase pre to post impacts in
brightness is much less obvious at 2.0-2.10 \microns than it is at 2.25-2.35
\micron, as shown in Figure \ref{fig_wlcompk_more}.  In the region of deep methane absorption centered at 2.3 \micron,
the planet became $\sim$2 times brighter after the impacts than it was before.  However, in the L-band there was no difference, even at wavelengths with deep methane absorption (See section 5.1).

\begin{figure}
\centering
\includegraphics[width=6.4in]{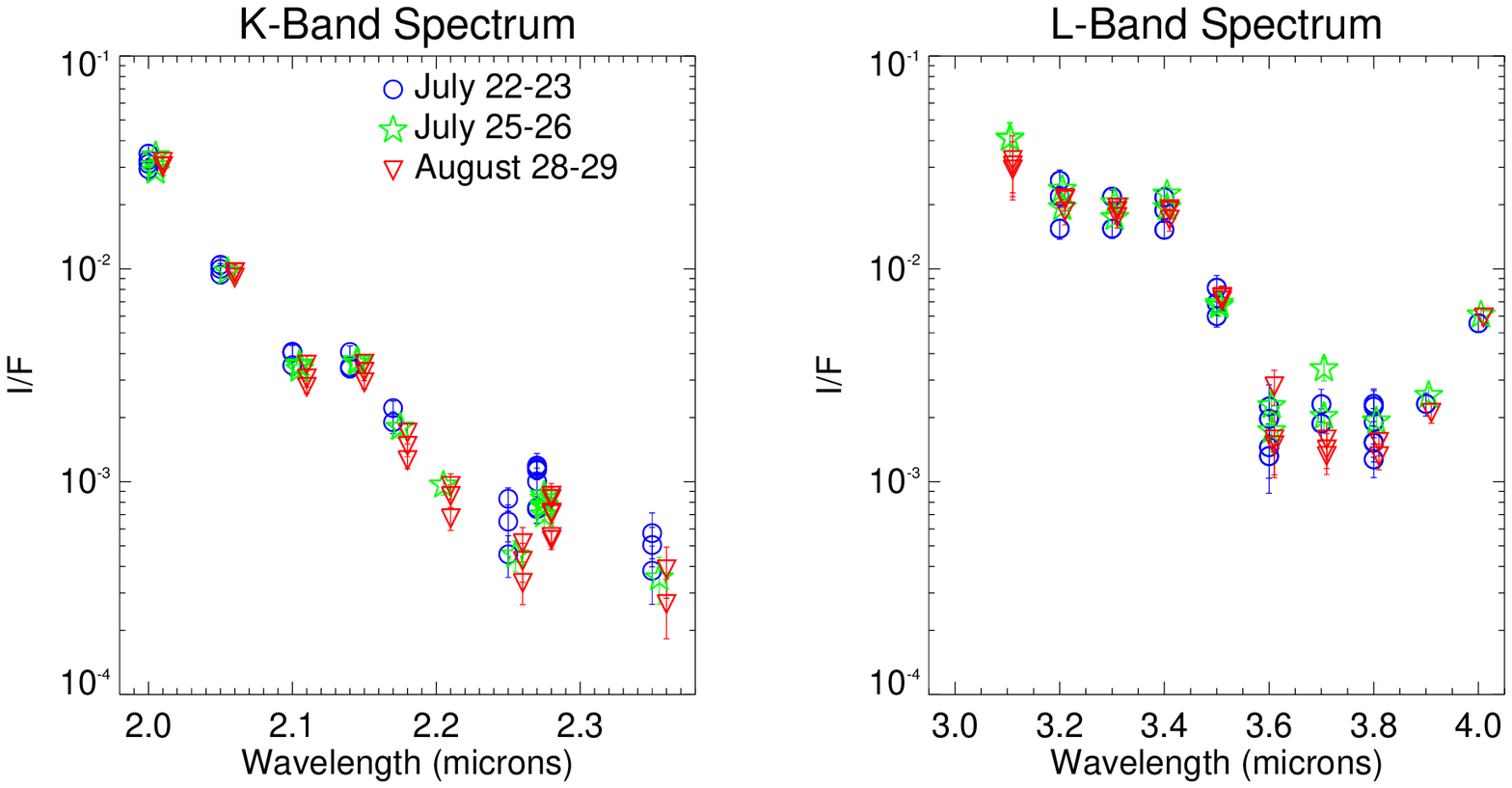}
\caption{$I/F$ spectra at K and L bands compared across three dates from July - August 1994.   There was very little overall change in brightness from July 23, the day after
 the last of the fragments impacted Jupiter, to August 29, the last day of our follow-up
observations.  Over time, winds
dispersed the material that was concentrated in the area of the impacts spots
in July. So in August, while the individual spots are not as bright, they
cover a larger area and the total flux is about the same.  The data
  from the different days is artificially off-set in wavelength for
  clarity. \label{fig_spectrum}}  
\end{figure}

\begin{figure}
\centering
\includegraphics[width=3.2in]{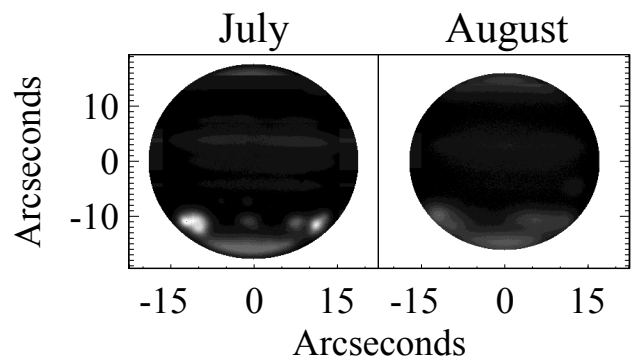}
\caption{Jupiter in late July versus late August.  Over time, winds
dispersed the material that was concentrated in the area of the impacts spots
in July. Jupiter appears smaller in the August figure as it was further away from Earth at the time.  The scaling in this image is the square root of the flux, so that the different latitudes can be more easily seen.\label{fig_julaug}}  
\end{figure}

Also of interest was how the total flux changed over the first month following the impacts.   Figure \ref{fig_spectrum}
shows the total brightness of the impacted planet with wavelength in the K-band and L-band compared across the photometric days.   There was very little overall change in brightness from July 23, the day after the last of the fragments impacted Jupiter, to August 29, the last day of our follow-up
observations. This was also noted in \citet{ban96}. Over time, winds
dispersed the material that was concentrated in the area of the impacts spots
in July. So in August, while the individual spots are not as bright, they
cover a larger area and the total flux is about the same, as shown in Figure \ref{fig_julaug}. Evidently the
settling time of the particulates exceeded a month.  Thus an observer of
Jupiter as an exoplanet would see a brightening of the planet over the
week-long timescale of the impacts followed by months of constant to very
slowly decreasing brightness.

\begin{figure}
\centering
\includegraphics[width=6.4in]{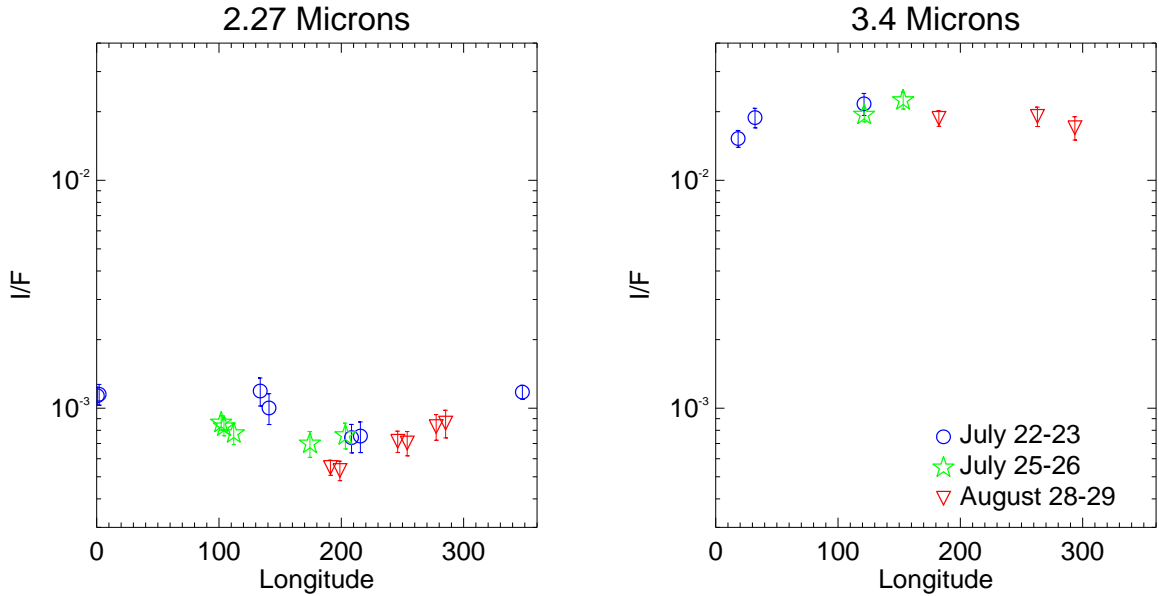}
\caption{Longitude dependence of $I/F$ at 2.276 \micron\ and at 3.4 \micron.  The difference in flux is due to the fact that the impacts, as well as other atmospheric features such as clouds and storms, were not homogeneously distributed across different longitudes. \label{fig_longitude}}  
\end{figure}

\begin{figure}
\centering
\includegraphics[width=3.2in]{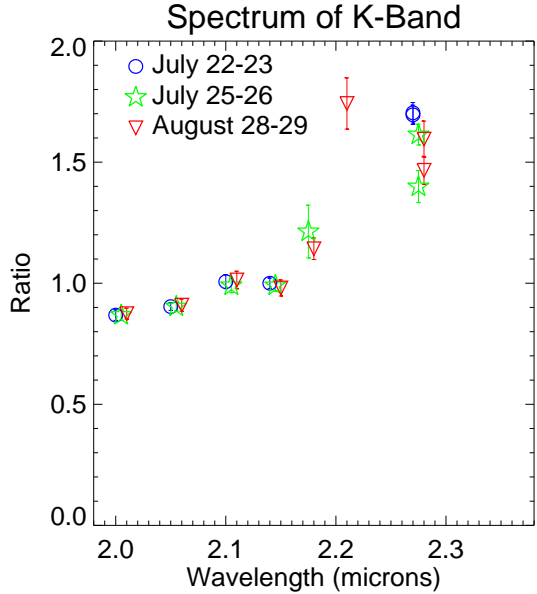}
\caption{The ratio of $I/F$ between the model planets with
  impact spots to the model planets without impact spots for images with central longitudes between 165$\degree$ and 225$\degree$, comparing $I/F$ from July and August. The data
  from the different days is artificially off-set in wavelength for
  clarity.  \label{fig_limitedspectrum}}  
\end{figure}

One complication in deriving a global behavior from our data is that each image sequence covers  a different range of Jovian
  longitudes. Since the impact spots were not homogeneously distributed on the
  planet, images centered at different longitudes have different flux as shown in Figure \ref{fig_longitude}.  This is also the cause in the difference in $I/F$ within one night for a given wavelength, as Jupiter rotates very quickly, with a rotation period of just under 10 hours \citep[and references therein]{gui04}. To examine the variability of the planet with time after
  the impacts, we use only data taken at longitudes between 165$\degree$ and 225$\degree$ to get a more direct comparison.  These longitudes were chosen because there were data from both months in the K-band.  Figure  \ref{fig_limitedspectrum} shows the ratio of the model planets with impact spots to those without impact spots from only those longitudes, and the effect is more obvious.   However, the observations in the L-band had no overlap in longitude for the photometric days, as shown by Figure \ref{fig_longitude}.  This made it impossible to do a direct comparison between days.

\section{Discussion and Conclusions}
\subsection{Analysis and Explanation of Results}
At the wavelengths where methane has strong absorption features, Jupiter's atmosphere has very high opacity. Results show that the typical atmospheric pressure level
probed by K-band observations drops from a few hundred mbar outside the 2.3 \micron\ methane band to $\sim$40 mbar inside it \citep{ban96}.  The SL9 impact spots, which are probably composed of sub-micron aerosols \citep{wes95}, are bright in the near-infrared because they efficiently reflect sunlight before it reaches altitudes where it can be absorbed by methane.  Their near-infrared brightness is comparable to that generated by the polar hazes, which are composed of hydrocarbon aerosols, i.e. submicron particles, generated in Jupiter's atmosphere by energetic particles launched off Io and carried to Jupiter along its magnetic field lines \citep{won00, cla04}.  In general, it is not known if exoplanets will have bright poles due to high altitude hazes, such that viewing angle will also affect the contrast between comet-deposited material and intrinsic haze.  After a month, particles smaller than 0.1 \micron\ would not have settled below 40 mbar while particles smaller than 0.5 \micron\ would not settle below 100 mbar \citep{wes95}.

\begin{figure}
\centering
\includegraphics[width=6.4in]{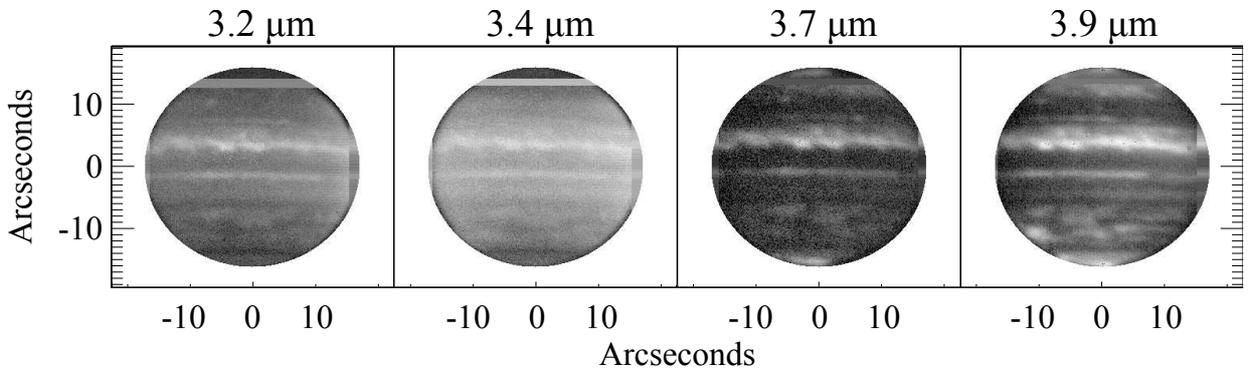}
\caption{Reproductions of Jupiter from August 29 at 3.2, 3.4, 3.7, and 3.9  \micron.  At these wave lengths, the central latitudes are much brighter, thus making the spots undetectable, even in the methane band. \label{fig_wlcompl_more}}  
\end{figure}

Although Galileo data shows that the 3.3 \micron\ methane
band also probes only the very uppermost layers of the atmosphere $\sim$1 mbar, \citep{mea95}
the L-band images of Jupiter show that there are bright bands of emission that do not correspond to the reflective areas at K-band.  Figure \ref{fig_wlcompl_more} shows no difference between the model planet images at that wavelength, so we conclude that there is thermal emission leaking out, comparable in magnitude to the scattered light off the impact spots.  

\subsection{Imaging Exoplanets}
At present, the few directly imaged exoplanets known (e.g. HR 8799 bcde
\citep{mar08}) are young, warm bodies glowing in their own thermal
emission. Although these planets are embedded in massive debris disks, their internal emission would likely wash out the effects of
impacting planetesimals. Current extreme adaptive optics
systems designed for planet-finding, such as GPI \citep{mac14} and SPHERE \citep{beu10} should directly
image young luminous planets. The ability to detect cool planets, such as
Jupiter with a surface temperature of $\sim$130$\degree$ K \citep{kuz76}, is probably left to future giant telescopes and high contrast space missions.   Proposed coronagraphs that can achieve a contrast ratio of better than $10^{-9}$ will be able to image these planets with separations of greater than 0.2\arcsec  \citep{law13}.  Atmospheric properties of warm planets are determined by measuring their spectra  at multiple wavelengths, and the presence of methane and other molecules are used to determine their temperatures and compositions.  This technique will not change in the
era when cooler planets can be observed in scattered light.   Already, there are $\sim$10 planets detected with the radial velocity method around main sequence stars with ages comparable to that of the Sun that are also amenable to high contrast imaging observations. More will be found as RV surveys continue for longer durations, and these will be the set of planets most amenable to direct imaging with future instruments.  (W. Traub, personal communications)

\subsection{Planetesimal Impact Detection}

Detecting these planetesimal impacts is important for several reasons.  The effect  of planetesimal impact spots would be a source of signal and a source of obscuration.    Because the planetesimal does not affect every longitude equally, bright spots that last many months  could allow the rotation rate of an exoplanet to be determined using the photometric variations as a function of longitude.  However, as accurate spectra are necessary to learn important information about planets, anything that might alter  the spectra has to be taken into account, and  given the length of time (i.e. many months) it takes the spots to fade, reflected light from the impact spots could obscure the molecular absorption bands normally used to determine atmospheric composition. The equivalent widths of molecular absorption lines will be lower than expected if high altitude clouds conceal the main sources of atmospheric opacity.  Additionally, high impact rates could affect the atmospheric composition of giant planets over time, both due to the energy released  by the impactors and by introducing, and keeping in non-equilibrium, molecular species that would not be found in equilibrium \citep{tur14}. For example, Jupiter has water in its stratosphere, i.e. above its water condensation level, only because of the SL9 impact \citep{har04,cav13}. While detailed spectroscopy of extrasolar planets such as that of solar system planets will never be feasible, knowing the cometary impact rate would allow realistic models of the evolution of an exoplanet's atmosphere. 


To show the same effect that SL9 caused on Jupiter, an exoplanet would need to show deep methane absorption. Methane has already been found in some hot Jupiters \citep[e.g.][]{Swain09}.  And fortunately, it seems likely that cooler exoplanets will also have methane.  All of the outer Solar System planets have higher methane mixing ratios than Jupiter \citep{kar94,pri84,Atreya1991,bai94}.  As these are the coldest planets imaged, with temperatures down to $\sim$65$\degree$ K for Neptune \citep{ing95}, methane could be the primary source of opacity for any exoplanet cooler than Jupiter.  And models indicate that there is significant methane absorption in the K-band  in planets with effective temperatures up to at least ~800$\degree$ K \citep{sau12}.  Even if there are cold exoplanets where methane is not a source of opacity, they are likely to have other molecular species, such as carbon dioxide or ammonia; comparable measurements could be made at wavelengths where those molecules have absorption bands. 

A planet covered in haze due to abundant aerosols would have less detectable comet impacts because it has less detectable atmospheric absorption lines.  Aerosol formation is not well understood and is certainly related to poorly understood vertical mixing that either does or does not bring methane above the condensation pressure.  Essentially what the impact does is mimic the reflective effect of such aerosols. When the depths of cool exoplanet molecular lines are eventually measured, comet impacts will be a source of particulates that could mimic large-coverage hazes. Looking at atmospheres over several years, to see if the particulates settle out or not, and determining if the planets sit in systems with debris disks can help distinguish the source of the high-altitude particulates.  

  For a planet that was impacted as rarely as Jupiter is now,  there is a low probability of observing the planet during the time when the impact spots were
still visible.  Planetesimals as small as 200 m could leave long-standing atmospheric debris, which would make detections possible, and those impact more frequently --- once or twice per decade in the case of Jupiter \citep{hue13}.   But a planet being impacted very frequently would also be problematic, as a planet that was impacted at the rate $\sim$1 per month would show weaker molecular 
absorption bands than one impacted rarely.

How often a planet is impacted by planetesimals is intimately related to the architecture of the planetary system because the small body populations and planets evolve together. Generally speaking, planets interior to massive planetesimal belts, planets close to their stars, and planets with larger gravitational cross sections are more likely to be impacted.

Other planetary systems could have significantly different impact rates. All
other things being equal, the number of comet impacts should be proportional
to the number of comets \citep{nak98}.  Planetesimals in other systems cannot
be detected directly, but the debris dust generated from their collisions
can be.  The number of small dust grains in a debris disk is measured based on
the IR excess in the system \citep[e.g.][]{bac92}.  Because the lifetime
of dust is relatively short, largely due to stellar radiation, the presence of dust
implies that there must be planetesimal collisions creating that dust.  To go
from the small dust to an estimate of the number of planetesimals is model
dependent because the intrinsic size distribution of bodies depends on a
number of factors including the bodies' size and strength
\citep[e.g.][]{wya11,gas13}. However, systems with more dust must have more
planetesimals. 

Cold debris disks appear to be approximately equally common for planet-hosting
and non-planet hosting stars \citep{bry09,eir13}, although the known
planets are all closer to their star than Jupiter is to the Sun.  Detectable
disks are typically 10--100 $\times$ dustier than our Edgeworth-Kuiper Belt
\citep{bry06,eir13}, with correspondingly larger parent body populations. We
would expect planets in these systems to be subject to frequent impacts.

Debris disks decay over time due to collisions that erode the parent
planetesimals and radiation effects that remove the small dust
grains. Therefore, impact rates are likely higher in younger systems. However
\citet{gas13} shows that even 10\%  of star of age similar to
the Sun have disks substantially more massive than our Edgeworth-Kuiper Belt.

Solar system architecture also has a large effect on the number of comets.  In
our Solar System, the comet supply has been severely depleted by giant planet
migration \citep{mey07}.  But that may not be the case in all other systems.
The ratio of comets to asteroids among planetesimals is a
function of how far the stable planetesimal belts are from their stars. In our Solar System,
asteroids are from the closer and warmer Asteroid Belt, while comets come
mainly from the Edgeworth-Kuiper Belt and the Oort Cloud, because those are at
temperatures at which volatiles - a defining component of comets- freeze
\citep{whi76}.  The disks that we know of around other stars are largely cold
and analogous to the Edgeworth-Kuiper Belt, and therefore capable of delivering impacts
to cold giant planets. Oort Cloud equivalents are not yet detectable.
A more complex factor is the planetary system as a whole.
Other planets can increase or decrease the number of cometary impacts
depending on their sizes and locations, but without detailed information on
specific systems, it's not possible to estimate the effects of different architecture on the impact rate.

There are many other things that could complicate the detection of a planetesimal impact through this method. For a planet much hotter than Jupiter, the
thermal emission would dominate the K-band, like it does the L-band with Jupiter, thus making the effect of a planetesimal impact much harder to detect.   A large asteroid would likely result in similar results as a comet impact. An impact on Jupiter in July 2009 was by what is believed to be an asteroid under 1 Km in diameter, yet the effects on the planet were thought to be relatively similar to that of SL9, especially at first \citep{ham10, san10, lis10}.    And while rotation rate could be detected, an unusual weather system, like the Great Red Spot on Jupiter, would have a similar effect.

\section*{Acknowledgments}

The Palomar data were taken in collaboration with Phil Nicholson and Gerry
Neugebauer, both of whom were essential to their quality and utility even
after 20 years. Gerry Neugebauer died as this paper was being prepared for
publication, but his long commitment to infrared astronomy generally and to
this work specifically is gratefully acknowledged.  This effort was supported
in part by the NASA Astrobiology Institute under cooperative agreement
NNA09DA81A. We would also like to thank the anonymous referees for their helpful comments.


\bibliographystyle{elsarticle-harv}
\bibliography{ref}

\begin{thebibliography}{49}
\expandafter\ifx\csname natexlab\endcsname\relax\def\natexlab#1{#1}\fi
\expandafter\ifx\csname url\endcsname\relax
  \def\url#1{\texttt{#1}}\fi
\expandafter\ifx\csname urlprefix\endcsname\relax\def\urlprefix{URL }\fi

\bibitem[{{Atreya} et~al.(1991){Atreya}, {Sandel}, and {Romani}}]{Atreya1991}
{Atreya}, S.~K., {Sandel}, B.~R., {Romani}, P.~N., 1991. {Photochemistry and
  vertical mixing}. In: {Bergstralh}, J.~T., {Miner}, E.~D., {Matthews}, M.~S.
  (Eds.), Uranus. University of Arizona Press, pp. 110--146.

\bibitem[{{Backman} et~al.(1992){Backman}, {Witteborn}, and {Gillett}}]{bac92}
{Backman}, D.~E., {Witteborn}, F.~C., {Gillett}, F.~C., Feb. 1992. {Infrared
  observations and thermal models of the Beta Pictoris disk}. \apj 385,
  670--679.

\bibitem[{{Baines} and {Hammel}(1994)}]{bai94}
{Baines}, K.~H., {Hammel}, H.~B., May 1994. {Clouds, hazes, and the
  stratospheric methane abundance in Neptune}. \icarus 109, 20--39.

\bibitem[{{Banfield} et~al.(1996){Banfield}, {Gierasch}, {Squyres},
  {Nicholson}, {Conrath}, and {Matthews}}]{ban96}
{Banfield}, D., {Gierasch}, P.~J., {Squyres}, S.~W., {Nicholson}, P.~D.,
  {Conrath}, B.~J., {Matthews}, K., Jun. 1996. {2 {$\mu$}m Spectophotometry of
  Jovian Stratospheric Aerosols --- Scattering Opacities, Vertical
  Distributions, and Wind Speeds}. \icarus 121, 389--410.

\bibitem[{{Beuzit} et~al.(2010){Beuzit}, {Boccaletti}, {Feldt}, {Dohlen},
  {Mouillet}, {Puget}, {Wildi}, {Abe}, {Antichi}, {Baruffolo}, {Baudoz},
  {Carbillet}, {Charton}, {Claudi}, {Desidera}, {Downing}, {Fabron},
  {Feautrier}, {Fedrigo}, {Fusco}, {Gach}, {Giro}, {Gratton}, {Henning},
  {Hubin}, {Joos}, {Kasper}, {Lagrange}, {Langlois}, {Lenzen}, {Moutou},
  {Pavlov}, {Petit}, {Pragt}, {Rabou}, {Rigal}, {Rochat}, {Roelfsema},
  {Rousset}, {Saisse}, {Schmid}, {Stadler}, {Thalmann}, {Turatto}, {Udry},
  {Vakili}, {Vigan}, and {Waters}}]{beu10}
{Beuzit}, J.-L., {Boccaletti}, A., {Feldt}, M., {Dohlen}, K., {Mouillet}, D.,
  {Puget}, P., {Wildi}, F., {Abe}, L., {Antichi}, J., {Baruffolo}, A.,
  {Baudoz}, P., {Carbillet}, M., {Charton}, J., {Claudi}, R., {Desidera}, S.,
  {Downing}, M., {Fabron}, C., {Feautrier}, P., {Fedrigo}, E., {Fusco}, T.,
  {Gach}, J.-L., {Giro}, E., {Gratton}, R., {Henning}, T., {Hubin}, N., {Joos},
  F., {Kasper}, M., {Lagrange}, A.-M., {Langlois}, M., {Lenzen}, R., {Moutou},
  C., {Pavlov}, A., {Petit}, C., {Pragt}, J., {Rabou}, P., {Rigal}, F.,
  {Rochat}, S., {Roelfsema}, R., {Rousset}, G., {Saisse}, M., {Schmid}, H.-M.,
  {Stadler}, E., {Thalmann}, C., {Turatto}, M., {Udry}, S., {Vakili}, F.,
  {Vigan}, A., {Waters}, R., Oct. 2010. {Direct Detection of Giant Extrasolar
  Planets with SPHERE on the VLT}. In: {Coud{\'e} Du Foresto}, V., {Gelino},
  D.~M., {Ribas}, I. (Eds.), Pathways Towards Habitable Planets. Vol. 430 of
  Astronomical Society of the Pacific Conference Series. p. 231.

\bibitem[{{Bryden} et~al.(2009){Bryden}, {Beichman}, {Carpenter}, {Rieke},
  {Stapelfeldt}, {Werner}, {Tanner}, {Lawler}, {Wyatt}, {Trilling}, {Su},
  {Blaylock}, and {Stansberry}}]{bry09}
{Bryden}, G., {Beichman}, C.~A., {Carpenter}, J.~M., {Rieke}, G.~H.,
  {Stapelfeldt}, K.~R., {Werner}, M.~W., {Tanner}, A.~M., {Lawler}, S.~M.,
  {Wyatt}, M.~C., {Trilling}, D.~E., {Su}, K.~Y.~L., {Blaylock}, M.,
  {Stansberry}, J.~A., Nov. 2009. {Planets and Debris Disks: Results from a
  Spitzer/MIPS Search for Infrared Excess}. \apj 705, 1226--1236.

\bibitem[{{Bryden} et~al.(2006){Bryden}, {Beichman}, {Trilling}, {Rieke},
  {Holmes}, {Lawler}, {Stapelfeldt}, {Werner}, {Gautier}, {Blaylock}, {Gordon},
  {Stansberry}, and {Su}}]{bry06}
{Bryden}, G., {Beichman}, C.~A., {Trilling}, D.~E., {Rieke}, G.~H., {Holmes},
  E.~K., {Lawler}, S.~M., {Stapelfeldt}, K.~R., {Werner}, M.~W., {Gautier},
  T.~N., {Blaylock}, M., {Gordon}, K.~D., {Stansberry}, J.~A., {Su}, K.~Y.~L.,
  Jan. 2006. {Frequency of Debris Disks around Solar-Type Stars: First Results
  from a Spitzer MIPS Survey}. \apj 636, 1098--1113.

\bibitem[{{Cavali{\'e}} et~al.(2013){Cavali{\'e}}, {Feuchtgruber}, {Lellouch},
  {de Val-Borro}, {Jarchow}, {Moreno}, {Hartogh}, {Orton}, {Greathouse},
  {Billebaud}, {Dobrijevic}, {Lara}, {Gonz{\'a}lez}, and {Sagawa}}]{cav13}
{Cavali{\'e}}, T., {Feuchtgruber}, H., {Lellouch}, E., {de Val-Borro}, M.,
  {Jarchow}, C., {Moreno}, R., {Hartogh}, P., {Orton}, G., {Greathouse}, T.~K.,
  {Billebaud}, F., {Dobrijevic}, M., {Lara}, L.~M., {Gonz{\'a}lez}, A.,
  {Sagawa}, H., May 2013. {Spatial distribution of water in the stratosphere of
  Jupiter from Herschel HIFI and PACS observations}. \aap 553, A21.

\bibitem[{{Clarke} et~al.(2004){Clarke}, {Grodent}, {Cowley}, {Bunce}, {Zarka},
  {Connerney}, and {Satoh}}]{cla04}
{Clarke}, J.~T., {Grodent}, D., {Cowley}, S.~W.~H., {Bunce}, E.~J., {Zarka},
  P., {Connerney}, J.~E.~P., {Satoh}, T., 2004. {Jupiter's aurora}. In:
  {Bagenal}, F., {Dowling}, T.~E., {McKinnon}, W.~B. (Eds.), Jupiter.~The
  Planet, Satellites and Magnetosphere. Cambridge University Press, Ch.~25, pp.
  639--670.

\bibitem[{{Crawford}(1997)}]{cra97}
{Crawford}, D.~A., May 1997. {Comet Shoemaker-Levy 9 Fragment Size Estimates:
  How Big was the Parent Body?} Annals of the New York Academy of Sciences 822,
  155.

\bibitem[{{Eiroa} et~al.(2013){Eiroa}, {Marshall}, {Mora}, {Montesinos},
  {Absil}, {Augereau}, {Bayo}, {Bryden}, {Danchi}, {del Burgo}, {Ertel},
  {Fridlund}, {Heras}, {Krivov}, {Launhardt}, {Liseau}, {L{\"o}hne},
  {Maldonado}, {Pilbratt}, {Roberge}, {Rodmann}, {Sanz-Forcada}, {Solano},
  {Stapelfeldt}, {Th{\'e}bault}, {Wolf}, {Ardila}, {Ar{\'e}valo}, {Beichmann},
  {Faramaz}, {Gonz{\'a}lez-Garc{\'{\i}}a}, {Guti{\'e}rrez}, {Lebreton},
  {Mart{\'{\i}}nez-Arn{\'a}iz}, {Meeus}, {Montes}, {Olofsson}, {Su}, {White},
  {Barrado}, {Fukagawa}, {Gr{\"u}n}, {Kamp}, {Lorente}, {Morbidelli},
  {M{\"u}ller}, {Mutschke}, {Nakagawa}, {Ribas}, and {Walker}}]{eir13}
{Eiroa}, C., {Marshall}, J.~P., {Mora}, A., {Montesinos}, B., {Absil}, O.,
  {Augereau}, J.~C., {Bayo}, A., {Bryden}, G., {Danchi}, W., {del Burgo}, C.,
  {Ertel}, S., {Fridlund}, M., {Heras}, A.~M., {Krivov}, A.~V., {Launhardt},
  R., {Liseau}, R., {L{\"o}hne}, T., {Maldonado}, J., {Pilbratt}, G.~L.,
  {Roberge}, A., {Rodmann}, J., {Sanz-Forcada}, J., {Solano}, E.,
  {Stapelfeldt}, K., {Th{\'e}bault}, P., {Wolf}, S., {Ardila}, D.,
  {Ar{\'e}valo}, M., {Beichmann}, C., {Faramaz}, V.,
  {Gonz{\'a}lez-Garc{\'{\i}}a}, B.~M., {Guti{\'e}rrez}, R., {Lebreton}, J.,
  {Mart{\'{\i}}nez-Arn{\'a}iz}, R., {Meeus}, G., {Montes}, D., {Olofsson}, G.,
  {Su}, K.~Y.~L., {White}, G.~J., {Barrado}, D., {Fukagawa}, M., {Gr{\"u}n},
  E., {Kamp}, I., {Lorente}, R., {Morbidelli}, A., {M{\"u}ller}, S.,
  {Mutschke}, H., {Nakagawa}, T., {Ribas}, I., {Walker}, H., Jul. 2013. {Dust
  around NEarby Stars. The survey observational results}. \aap 555, A11.

\bibitem[{{Elias} et~al.(1982){Elias}, {Frogel}, {Matthews}, and
  {Neugebauer}}]{eli82}
{Elias}, J.~H., {Frogel}, J.~A., {Matthews}, K., {Neugebauer}, G., Jul. 1982.
  {Infrared standard stars}. \aj 87, 1029--1034.

\bibitem[{{G{\'a}sp{\'a}r} et~al.(2013){G{\'a}sp{\'a}r}, {Rieke}, and
  {Balog}}]{gas13}
{G{\'a}sp{\'a}r}, A., {Rieke}, G.~H., {Balog}, Z., May 2013. {The Collisional
  Evolution of Debris Disks}. \apj 768, 25.

\bibitem[{{Giorgini} et~al.(1996){Giorgini}, {Yeomans}, {Chamberlin}, {Chodas},
  {Jacobson}, {Keesey}, {Lieske}, {Ostro}, {Standish}, and {Wimberly}}]{gio96}
{Giorgini}, J.~D., {Yeomans}, D.~K., {Chamberlin}, A.~B., {Chodas}, P.~W.,
  {Jacobson}, R.~A., {Keesey}, M.~S., {Lieske}, J.~H., {Ostro}, S.~J.,
  {Standish}, E.~M., {Wimberly}, R.~N., Sep. 1996. {JPL's On-Line Solar System
  Data Service}. In: AAS/Division for Planetary Sciences Meeting Abstracts
  \#28. Vol.~28 of Bulletin of the American Astronomical Society. p. 1158.

\bibitem[{{Guillot} et~al.(2004){Guillot}, {Stevenson}, {Hubbard}, and
  {Saumon}}]{gui04}
{Guillot}, T., {Stevenson}, D., {Hubbard}, W., {Saumon}, D., 2004. {Jupiter's
  Interior}. In: {Bagenal}, F., {Dowling}, T.~E., {McKinnon}, W.~B. (Eds.),
  Jupiter.~The Planet, Satellites and Magnetosphere. Cambridge University
  Press, Ch.~2, pp. 35--57.

\bibitem[{{Hammel} et~al.(1995){Hammel}, {Beebe}, {Ingersoll}, {Orton},
  {Mills}, {Simon}, {Chodas}, {Clarke}, {de Jong}, {Dowling}, {Harrington},
  {Huber}, {Karkoschka}, {Santori}, {Toigo}, {Yeomans}, and {West}}]{ham95}
{Hammel}, H.~B., {Beebe}, R.~F., {Ingersoll}, A.~P., {Orton}, G.~S., {Mills},
  J.~R., {Simon}, A.~A., {Chodas}, P., {Clarke}, J.~T., {de Jong}, E.,
  {Dowling}, T.~E., {Harrington}, J., {Huber}, L.~F., {Karkoschka}, E.,
  {Santori}, C.~M., {Toigo}, A., {Yeomans}, D., {West}, R.~A., Mar. 1995. {HST
  Imaging of Atmospheric Phenomena Created by the Impact of Comet
  Shoemaker-Levy 9}. Science 267, 1288--1296.

\bibitem[{{Hammel} et~al.(2010){Hammel}, {Wong}, {Clarke}, {de Pater},
  {Fletcher}, {Hueso}, {Noll}, {Orton}, {P{\'e}rez-Hoyos},
  {S{\'a}nchez-Lavega}, {Simon-Miller}, and {Yanamandra-Fisher}}]{ham10}
{Hammel}, H.~B., {Wong}, M.~H., {Clarke}, J.~T., {de Pater}, I., {Fletcher},
  L.~N., {Hueso}, R., {Noll}, K., {Orton}, G.~S., {P{\'e}rez-Hoyos}, S.,
  {S{\'a}nchez-Lavega}, A., {Simon-Miller}, A.~A., {Yanamandra-Fisher}, P.~A.,
  Jun. 2010. {Jupiter After the 2009 Impact: Hubble Space Telescope Imaging of
  the Impact-generated Debris and its Temporal Evolution}. \apjl 715,
  L150--L154.

\bibitem[{{Harrington} et~al.(2004){Harrington}, {de Pater}, {Brecht},
  {Demind}, {Meadows}, {Zahnle}, and {Nicholson}}]{har04}
{Harrington}, J., {de Pater}, I., {Brecht}, S., {Demind}, D., {Meadows}, V.,
  {Zahnle}, K., {Nicholson}, P., 2004. {Lessons from Shoemaker-Levy 9 about
  Jupiter and Planetary Impacts}. In: {Bagenal}, F., {Dowling}, T.~E.,
  {McKinnon}, W.~B. (Eds.), Jupiter.~The Planet, Satellites and Magnetosphere.
  Cambridge University Press, Ch.~7, pp. 159--184.

\bibitem[{{Hueso} et~al.(2013){Hueso}, {P{\'e}rez-Hoyos}, {S{\'a}nchez-Lavega},
  {Wesley}, {Hall}, {Go}, {Tachikawa}, {Aoki}, {Ichimaru}, {Pond},
  {Korycansky}, {Palotai}, {Chappell}, {Rebeli}, {Harrington}, {Delcroix},
  {Wong}, {de Pater}, {Fletcher}, {Hammel}, {Orton}, {Tabe}, {Watanabe}, and
  {Moreno}}]{hue13}
{Hueso}, R., {P{\'e}rez-Hoyos}, S., {S{\'a}nchez-Lavega}, A., {Wesley}, A.,
  {Hall}, G., {Go}, C., {Tachikawa}, M., {Aoki}, K., {Ichimaru}, M., {Pond},
  J.~W.~T., {Korycansky}, D.~G., {Palotai}, C., {Chappell}, G., {Rebeli}, N.,
  {Harrington}, J., {Delcroix}, M., {Wong}, M., {de Pater}, I., {Fletcher},
  L.~N., {Hammel}, H., {Orton}, G.~S., {Tabe}, I., {Watanabe}, J., {Moreno},
  J.~C., Dec. 2013. {Impact flux on Jupiter: From superbolides to large-scale
  collisions}. \aap 560, A55.

\bibitem[{{Ingersoll} et~al.(1995){Ingersoll}, {Barnet}, {Beebe}, {Flasar},
  {Hinson}, {Limaye}, {Sromovsky}, and {Suomi}}]{ing95}
{Ingersoll}, A.~P., {Barnet}, C.~D., {Beebe}, R.~F., {Flasar}, F.~M., {Hinson},
  D.~P., {Limaye}, S.~S., {Sromovsky}, L.~A., {Suomi}, V.~E., 1995. {Dynamic
  meteorology of Neptune.} In: {Cruikshank}, D.~P., {Matthews}, M.~S.,
  {Schumann}, A.~M. (Eds.), Neptune and Triton. pp. 613--682.

\bibitem[{{Karkoschka}(1994)}]{kar94}
{Karkoschka}, E., Sep. 1994. {Spectrophotometry of the jovian planets and Titan
  at 300- to 1000-nm wavelength: The methane spectrum}. \icarus 111, 174--192.

\bibitem[{Kim et~al.(2010)Kim, Geballe, Kim, Jung, Seo, and Minh}]{kim10}
Kim, S.~J., Geballe, T., Kim, J., Jung, A., Seo, H., Minh, Y., 2010.
  High-resolution 3-Î¼m spectra of jupiter: Latitudinal spectral variations
  influenced by molecules, clouds, and haze. Icarus 208~(2), 837 -- 849.
\newline\urlprefix\url{http://www.sciencedirect.com/science/article/pii/S0019103510001338}

\bibitem[{{Klassen} and {Bell}(2003)}]{klassenbell2003}
{Klassen}, D.~R., {Bell}, J.~F., May 2003. {Radiance factor calibration of
  near-infrared spectral images of Mars}. \icarus 163, 66--77.

\bibitem[{{Kuzmin} and {Smirnova}(1976)}]{kuz76}
{Kuzmin}, A.~D., {Smirnova}, T.~V., Jun. 1976. {Temperature and pressure in the
  atmosphere of Jupiter}. \sovast 19, 756--758.

\bibitem[{{Lisse} et~al.(2010){Lisse}, {Orton}, {Yanamandra-Fisher},
  {Fletcher}, {Depater}, and {Hammel}}]{lis10}
{Lisse}, C., {Orton}, G., {Yanamandra-Fisher}, P., {Fletcher}, L., {Depater},
  I., {Hammel}, H., May 2010. {Spectroscopic Evidence for the Asteroidal Nature
  of the July 2009 Jovian Impactor}. In: EGU General Assembly Conference
  Abstracts. Vol.~12 of EGU General Assembly Conference Abstracts. p. 12186.

\bibitem[{{Macintosh} et~al.(2014){Macintosh}, {Graham}, {Ingraham},
  {Konopacky}, {Marois}, {Perrin}, {Poyneer}, {Bauman}, {Barman}, {Burrows},
  {Cardwell}, {Chilcote}, {De Rosa}, {Dillon}, {Doyon}, {Dunn}, {Erikson},
  {Fitzgerald}, {Gavel}, {Goodsell}, {Hartung}, {Hibon}, {Kalas}, {Larkin},
  {Maire}, {Marchis}, {Marley}, {McBride}, {Millar-Blanchaer}, {Morzinski},
  {Norton}, {Oppenheimer}, {Palmer}, {Patience}, {Pueyo}, {Rantakyro},
  {Sadakuni}, {Saddlemyer}, {Savransky}, {Serio}, {Soummer},
  {Sivaramakrishnan}, {Song}, {Thomas}, {Wallace}, {Wiktorowicz}, and
  {Wolff}}]{mac14}
{Macintosh}, B., {Graham}, J.~R., {Ingraham}, P., {Konopacky}, Q., {Marois},
  C., {Perrin}, M., {Poyneer}, L., {Bauman}, B., {Barman}, T., {Burrows}, A.,
  {Cardwell}, A., {Chilcote}, J., {De Rosa}, R.~J., {Dillon}, D., {Doyon}, R.,
  {Dunn}, J., {Erikson}, D., {Fitzgerald}, M., {Gavel}, D., {Goodsell}, S.,
  {Hartung}, M., {Hibon}, P., {Kalas}, P.~G., {Larkin}, J., {Maire}, J.,
  {Marchis}, F., {Marley}, M., {McBride}, J., {Millar-Blanchaer}, M.,
  {Morzinski}, K., {Norton}, A., {Oppenheimer}, B.~R., {Palmer}, D.,
  {Patience}, J., {Pueyo}, L., {Rantakyro}, F., {Sadakuni}, N., {Saddlemyer},
  L., {Savransky}, D., {Serio}, A., {Soummer}, R., {Sivaramakrishnan}, A.,
  {Song}, I., {Thomas}, S., {Wallace}, J.~K., {Wiktorowicz}, S., {Wolff}, S.,
  2014. {The Gemini Planet Imager: First Light}. Proc. Natl. Acad. Sci.

\bibitem[{{Marois} et~al.(2008){Marois}, {Macintosh}, {Barman}, {Zuckerman},
  {Song}, {Patience}, {Lafreni{\`e}re}, and {Doyon}}]{mar08}
{Marois}, C., {Macintosh}, B., {Barman}, T., {Zuckerman}, B., {Song}, I.,
  {Patience}, J., {Lafreni{\`e}re}, D., {Doyon}, R., Nov. 2008. {Direct Imaging
  of Multiple Planets Orbiting the Star HR 8799}. Science 322, 1348--.

\bibitem[{{Marshall} et~al.(2014){Marshall}, {Moro-Mart{\'{\i}}n}, {Eiroa},
  {Kennedy}, {Mora}, {Sibthorpe}, {Lestrade}, {Maldonado}, {Sanz-Forcada},
  {Wyatt}, {Matthews}, {Horner}, {Montesinos}, {Bryden}, {del Burgo},
  {Greaves}, {Ivison}, {Meeus}, {Olofsson}, {Pilbratt}, and
  {White}}]{marshall2014}
{Marshall}, J.~P., {Moro-Mart{\'{\i}}n}, A., {Eiroa}, C., {Kennedy}, G.,
  {Mora}, A., {Sibthorpe}, B., {Lestrade}, J.-F., {Maldonado}, J.,
  {Sanz-Forcada}, J., {Wyatt}, M.~C., {Matthews}, B., {Horner}, J.,
  {Montesinos}, B., {Bryden}, G., {del Burgo}, C., {Greaves}, J.~S., {Ivison},
  R.~J., {Meeus}, G., {Olofsson}, G., {Pilbratt}, G.~L., {White}, G.~J., May
  2014. {Correlations between the stellar, planetary, and debris components of
  exoplanet systems observed by Herschel}. \aap 565, A15.

\bibitem[{{Meadows} et~al.(1995){Meadows}, {Crisp}, {Orton}, {Brooke}, and
  {Spencer}}]{mea95}
{Meadows}, V., {Crisp}, D., {Orton}, G., {Brooke}, T., {Spencer}, J., 1995.
  {AAT IRIS observations of the SL-9 impacts and initial fireball evolution.}
  In: {West}, R.~M., {B{\"o}hnhardt}, H. (Eds.), European SL-9/Jupiter
  Workshop. Vol.~52 of European Southern Observatory Conference and Workshop
  Proceedings. pp. 129--134.

\bibitem[{{Meyer} et~al.(2007){Meyer}, {Backman}, {Weinberger}, and
  {Wyatt}}]{mey07}
{Meyer}, M.~R., {Backman}, D.~E., {Weinberger}, A.~J., {Wyatt}, M.~C., 2007.
  {Evolution of Circumstellar Disks Around Normal Stars: Placing Our Solar
  System in Context}. Protostars and Planets V, 573--588.

\bibitem[{{Morley} et~al.(2014){Morley}, {Marley}, {Fortney}, {Lupu}, {Saumon},
  {Greene}, and {Lodders}}]{mor14}
{Morley}, C.~V., {Marley}, M.~S., {Fortney}, J.~J., {Lupu}, R., {Saumon}, D.,
  {Greene}, T., {Lodders}, K., May 2014. {Water Clouds in Y Dwarfs and
  Exoplanets}. \apj 787, 78.

\bibitem[{{Nakamura} and {Kurahashi}(1998)}]{nak98}
{Nakamura}, T., {Kurahashi}, H., Feb. 1998. {Collisional probability of
  periodic comets with the terrestrial planets - an invalid case of analytic
  formulation}. \aj 115, 848.

\bibitem[{{Nicholson} et~al.(1995){Nicholson}, {Gierasch}, {Hayward}, {McGhee},
  {Moersch}, {Squyres}, {Van Cleve}, {Matthews}, {Neugebauer}, {Shupe},
  {Weinberger}, {Miles}, and {Conrath}}]{nic95}
{Nicholson}, P.~D., {Gierasch}, P.~J., {Hayward}, T.~L., {McGhee}, C.~A.,
  {Moersch}, J.~E., {Squyres}, S.~W., {Van Cleve}, J., {Matthews}, K.,
  {Neugebauer}, G., {Shupe}, D., {Weinberger}, A., {Miles}, J.~W., {Conrath},
  B.~J., 1995. {Palomar observations of the R impact of comet Shoemaker-Levy 9:
  I. Light curves}. \grl 22, 1613--1616.

\bibitem[{{Prinn} et~al.(1984){Prinn}, {Larson}, {Caldwell}, and
  {Gautier}}]{pri84}
{Prinn}, R.~G., {Larson}, H.~P., {Caldwell}, J.~J., {Gautier}, D., 1984.
  {Composition and chemistry of Saturn's atmosphere}. In: {Gehrels}, T.,
  {Matthews}, M.~S. (Eds.), Saturn. University of Arizona Press, pp. 88--149.

\bibitem[{{Sanchez-Lavega} et~al.(1998){Sanchez-Lavega}, {G{\'o}mez}, {Rojas},
  {Acarreta}, {Lecacheux}, {Colas}, {Hueso}, and {Arregui}}]{san98}
{Sanchez-Lavega}, A., {G{\'o}mez}, J.~M., {Rojas}, J.~F., {Acarreta}, J.~R.,
  {Lecacheux}, J., {Colas}, F., {Hueso}, R., {Arregui}, J., Feb. 1998.
  {Long-Term Evolution of Comet SL-9 Impact Features: July 1994-September
  1996}. \icarus 131, 341--357.

\bibitem[{{S{\'a}nchez-Lavega} et~al.(2010){S{\'a}nchez-Lavega}, {Wesley},
  {Orton}, {Hueso}, {Perez-Hoyos}, {Fletcher}, {Yanamandra-Fisher},
  {Legarreta}, {de Pater}, {Hammel}, {Simon-Miller}, {Gomez-Forrellad},
  {Ortiz}, {Garc{\'{\i}}a-Melendo}, {Puetter}, and {Chodas}}]{san10}
{S{\'a}nchez-Lavega}, A., {Wesley}, A., {Orton}, G., {Hueso}, R.,
  {Perez-Hoyos}, S., {Fletcher}, L.~N., {Yanamandra-Fisher}, P., {Legarreta},
  J., {de Pater}, I., {Hammel}, H., {Simon-Miller}, A., {Gomez-Forrellad},
  J.~M., {Ortiz}, J.~L., {Garc{\'{\i}}a-Melendo}, E., {Puetter}, R.~C.,
  {Chodas}, P., Jun. 2010. {The Impact of a Large Object on Jupiter in 2009
  July}. \apjl 715, L155--L159.

\bibitem[{{Saumon} et~al.(2012){Saumon}, {Marley}, {Abel}, {Frommhold}, and
  {Freedman}}]{sau12}
{Saumon}, D., {Marley}, M.~S., {Abel}, M., {Frommhold}, L., {Freedman}, R.~S.,
  May 2012. {New H$_{2}$ Collision-induced Absorption and NH$_{3}$ Opacity and
  the Spectra of the Coolest Brown Dwarfs}. \apj 750, 74.

\bibitem[{{Schenk} et~al.(2004){Schenk}, {Chapman}, {Zahnle}, and
  {Moore}}]{sch04}
{Schenk}, P.~M., {Chapman}, C.~R., {Zahnle}, K., {Moore}, J.~M., 2004. {Ages
  and interiors: the cratering record of the Galilean satellites}. In:
  {Bagenal}, F., {Dowling}, T.~E., {McKinnon}, W.~B. (Eds.), Jupiter.~The
  Planet, Satellites and Magnetosphere. Cambridge University Press, Ch.~17, pp.
  427--456.

\bibitem[{{Seidelmann} et~al.(2007){Seidelmann}, {Archinal}, {A'Hearn},
  {Conrad}, {Consolmagno}, {Hestroffer}, {Hilton}, {Krasinsky}, {Neumann},
  {Oberst}, {Stooke}, {Tedesco}, {Tholen}, {Thomas}, and {Williams}}]{sei07}
{Seidelmann}, P.~K., {Archinal}, B.~A., {A'Hearn}, M.~F., {Conrad}, A.,
  {Consolmagno}, G.~J., {Hestroffer}, D., {Hilton}, J.~L., {Krasinsky}, G.~A.,
  {Neumann}, G., {Oberst}, J., {Stooke}, P., {Tedesco}, E.~F., {Tholen}, D.~J.,
  {Thomas}, P.~C., {Williams}, I.~P., Jul. 2007. {Report of the IAU/IAG Working
  Group on cartographic coordinates and rotational elements: 2006}. Celestial
  Mechanics and Dynamical Astronomy 98, 155--180.

\bibitem[{Shaklan(2013)}]{law13}
Shaklan, S. (Ed.), 2013. Survey of experimental results in high-contrast
  imaging for future exoplanet missions. Vol. 8864.
\newline\urlprefix\url{http://dx.doi.org/10.1117/12.2021302}

\bibitem[{{Strom} et~al.(2005){Strom}, {Malhotra}, {Ito}, {Yoshida}, and
  {Kring}}]{str05}
{Strom}, R.~G., {Malhotra}, R., {Ito}, T., {Yoshida}, F., {Kring}, D.~A., Sep.
  2005. {The Origin of Planetary Impactors in the Inner Solar System}. Science
  309, 1847--1850.

\bibitem[{{Swain} et~al.(2009){Swain}, {Tinetti}, {Vasisht}, {Deroo},
  {Griffith}, {Bouwman}, {Chen}, {Yung}, {Burrows}, {Brown}, {Matthews},
  {Rowe}, {Kuschnig}, and {Angerhausen}}]{Swain09}
{Swain}, M.~R., {Tinetti}, G., {Vasisht}, G., {Deroo}, P., {Griffith}, C.,
  {Bouwman}, J., {Chen}, P., {Yung}, Y., {Burrows}, A., {Brown}, L.~R.,
  {Matthews}, J., {Rowe}, J.~F., {Kuschnig}, R., {Angerhausen}, D., Oct. 2009.
  {Water, Methane, and Carbon Dioxide Present in the Dayside Spectrum of the
  Exoplanet HD 209458b}. \apj 704, 1616--1621.

\bibitem[{{Taylor} and {Calcutt}(1984)}]{tay84}
{Taylor}, F.~W., {Calcutt}, S.~B., Dec. 1984. {Near-infrared spectroscopy of
  the atmosphere of Jupiter}. \jqsrt 32, 463--477.

\bibitem[{Turrini et~al.(2014)Turrini, Nelson, and Barbieri}]{tur14}
Turrini, D., Nelson, R., Barbieri, M., 2014. The role of planetary formation
  and evolution in shaping the composition of exoplanetary atmospheres.
  Experimental Astronomy, 1--22.
\newline\urlprefix\url{http://dx.doi.org/10.1007/s10686-014-9401-6}

\bibitem[{{West} et~al.(1995){West}, {Karkoschka}, {Friedson}, {Seymour},
  {Baines}, and {Hammel}}]{wes95}
{West}, R.~A., {Karkoschka}, E., {Friedson}, A.~J., {Seymour}, M., {Baines},
  K.~H., {Hammel}, H.~B., Mar. 1995. {Impact Debris Particles in Jupiter's
  Stratosphere}. Science 267, 1296--1301.

\bibitem[{{Whipple} and {Huebner}(1976)}]{whi76}
{Whipple}, F.~L., {Huebner}, W.~F., 1976. {Physical processes in comets}. \araa
  14, 143--172.

\bibitem[{{Wong} et~al.(2000){Wong}, {Lee}, {Yung}, and {Ajello}}]{won00}
{Wong}, A.-S., {Lee}, A.~Y.~T., {Yung}, Y.~L., {Ajello}, J.~M., May 2000.
  {Jupiter: Aerosol Chemistry in the Polar Atmosphere}. \apjl 534, L215--L217.

\bibitem[{{Wyatt} et~al.(2011){Wyatt}, {Clarke}, and {Booth}}]{wya11}
{Wyatt}, M.~C., {Clarke}, C.~J., {Booth}, M., Oct. 2011. {Debris disk size
  distributions: steady state collisional evolution with Poynting-Robertson
  drag and other loss processes}. Celestial Mechanics and Dynamical Astronomy
  111, 1--28.

\bibitem[{{Zahnle} et~al.(2003){Zahnle}, {Schenk}, {Levison}, and
  {Dones}}]{zah03}
{Zahnle}, K., {Schenk}, P., {Levison}, H., {Dones}, L., Jun. 2003. {Cratering
  rates in the outer Solar System}. \icarus 163, 263--289.

\end{thebibliography}


\end{document}